\documentclass[12pt]{article}
\usepackage{natbib}
\usepackage{amsmath}
\usepackage{bm} 
\usepackage{amssymb} 
\usepackage{mathrsfs} 
\usepackage{mathtools} 

\usepackage{graphicx}
\graphicspath{{figures/}} 
\usepackage{enumerate}
\usepackage{url} 
\usepackage{xr} 
\newcommand{\blind}{1}

\begin{document}

\def\spacingset#1{\renewcommand{\baselinestretch}%
{#1}\small\normalsize} \spacingset{1}


\if1\blind
{
 \title{\bf Estimating Causal Impacts of Scaling a Voluntary Policy Intervention}
 \author{Irina Degtiar\thanks{
  Contact: Dr. Irina Degtiar, Mathematica, idegtiar@mathematica-mpr.com. This work was supported by the U.S. Department of Health and Human Services, Centers for Medicare \& Medicaid Services under contracts HHSM-500-2014-00034I/HHSM-500-T0010 and HHSM-500-2014-00034I/75FCMC19F0005}\hspace{.2cm}\\
  Mathematica, Inc., Cambridge, MA, USA \\
  and \\
  Mariel Finucane \\
  Blue Cross Blue Shield of Massachusetts, Boston, MA, USA}
 \maketitle
} \fi

\if0\blind
{
 \bigskip
 \bigskip
 \bigskip
 \begin{center}
  {\LARGE\bf Estimating Causal Impacts of Scaling a Voluntary Policy Intervention}
\end{center}
 \medskip
} \fi

\bigskip
\begin{abstract}
	Evaluations often inform future program implementation decisions. However, the implementation context may differ, sometimes substantially, from the evaluation study context. This difference leads to uncertainty regarding the relevance of evaluation findings to future decisions. Voluntary interventions pose another challenge to generalizability, as we do not know precisely who will volunteer for the intervention in the future. We present a novel approach for estimating target population average treatment effects among the treated by generalizing results from an observational study to projected volunteers within the target population (the treated group). Our estimation approach can accommodate flexible outcome regression estimators such as Bayesian Additive Regression Trees (BART) and Bayesian Causal Forests (BCF). Our generalizability approach incorporates uncertainty regarding target population treatment status into the posterior credible intervals to better reflect the uncertainty of scaling a voluntary intervention. In a simulation based on real data, we demonstrate that these flexible estimators (BCF and BART) improve performance over estimators that rely on parametric regressions. We use our approach to estimate impacts of scaling up Comprehensive Primary Care Plus, a health care payment model intended to improve quality and efficiency of primary care, and we demonstrate the promise of scaling to a targeted subgroup of practices.
\end{abstract}

\noindent%
\vfill

\newpage
\spacingset{1.5} 
\section{Introduction}
\label{sec:intro}

Policy impact evaluations seek to guide future policy decisions, such as whether to scale up an intervention at the conclusion of the evaluation study. Much consideration is given to ensuring that impact estimates are internally valid and that all confounders have been measured and adjusted for. Less commonly considered is the external validity of the study, namely, the extent to which findings can hold for future contexts and populations. For example, from 2017 to 2021, the Centers for Medicare \& Medicaid Services (CMS) led the largest primary care payment and delivery reform model at the time, Comprehensive Primary Care Plus (CPC+), which aimed to improve quality, access, and efficiency of primary care \citep{swankoski2022}. Two versions (tracks) of CPC+ were offered within select regions in the United States, within which primary care practices could volunteer to participate. Both tracks had the same goals, but Track 2 had more stringent requirements than Track 1 and offered greater payment support. The evaluation study assessed the causal impacts of CPC+ on Medicare beneficiaries served by these volunteering practices, with the aim of supporting CMS' decision making about whether to expand the payment model beyond the original model participants. However, causal impacts of CPC+ in the evaluation sample of practices may differ from those in a nationwide scale-up if practices respond differently to the intervention---when effect modification exists---and if the distribution of those effect modifiers differs between the study population and target population of interest. 

Addressing the discrepancy between study and target populations to extend impact results beyond the evaluation study at hand requires generalizability and transportability methods. These methods have attracted increasing attention \citep{degtiar2023}, resulting in approaches that use outcome regressions such as ordinary least squares models or Bayesian Additive Regression Trees (BART) \citep{hill2011, green2012, kern2016}; propensity of selection weighting approaches like inverse probability of participation weighting (IPPW) \citep{cole2010,correa2018}; and double-robust estimators, such as the targeted maximum likelihood estimator \citep{rudolph2017} and augmented inverse probability of participation weighting (AIPPW) \citep{dahabreh2020}. However, most approaches, with the exception of BART, have relied on parametric modeling assumptions that relationships between covariates and the outcome are linear and additive. Few existing approaches allow for flexible modeling, which is particularly important when generalizing results from large observational studies with many confounders and effect modifiers, studies in which the true relationships between covariates and the outcome are unknown and not easily captured through simple linear additive relationships.

Estimating the impact of a CPC+ scale-up requires additional considerations novel to the generalizability literature because of the voluntary nature of CPC+ participation. Namely, because scale-up participation would likely remain voluntary, membership in the target treated population is uncertain: it is unclear which practices nationwide would volunteer for a scale-up. Although there is a large literature on scale-up implementation \citep{barker2016, powell2015, simmons2010}, and several approaches exist for estimating impacts of a policy model scale-up \citep{attanasio2003, flores2013, gechter2015}, to our knowledge, no literature addresses generalizability to a target treated population that is not enumerable (because of uncertainty about which practices would volunteer for the scaled-up intervention). Similar considerations arise when estimating the impacts of offering other voluntary interventions beyond the study setting, such as offering a novel math curriculum in new schools or a novel training program in new job centers.

To address these shortcomings, we present a novel generalizability approach for estimating target population average treatment effects among the treated (TATT). The approach can accommodate nonparametric outcome regression estimators such as BART and Bayesian Causal Forests (BCF) \citep{hahn2020} to extend inference from an observational or randomized study sample to the target treated population while accounting for confounding and effect heterogeneity in a data-driven fashion. As Bayesian estimators, BART and BCF methods allow for incorporating additional sources of uncertainty into the credible intervals, such as uncertainty about what factors will drive participation in the scaled-up intervention. Incorporating these sources of uncertainty into the final Bayesian credible intervals helps avoid overstating confidence in estimated scale-up effects. Thus, point estimates and uncertainty bounds directly capture uncertainty about which members of the target population will volunteer for a scale-up.

The TATT generalizability estimator first estimates a propensity to volunteer for all units in the \textit{target sample}, the sample to which the scaled-up intervention will be offered, then estimates impacts for all target sample units from a regression fit to the evaluation sample using BART, BCF, or another estimator. Impacts for the \textit{target treated sample} of volunteers consist of propensity-for-volunteering weighted averages of target sample impact estimates. We apply this novel generalizability estimator to estimate impacts of scaling CPC+ nationwide.

The remainder of this article is organized as follows. Section \ref{sec:notation} introduces notation, the estimand, assumptions, and identification. Section \ref{sec:estimation} presents our novel approach for estimating the TATT and introduces BART, BCF, and other estimator choices. Section \ref{sec:sim} assesses our generalizability estimator's performance in a simulation study based on real data. Section \ref{sec:cpcplus} estimates the impact of scaling CPC+ nationwide. We conclude with a discussion in Section \ref{sec:discussion}.

\section{Notation, Assumptions, and Causal Quantities}\label{sec:notation}
\subsection{Data Structure, Notation, and Estimand}
\begin{figure} 
	\spacingset{1}
	\begin{center}
		\includegraphics*[width=\textwidth, keepaspectratio=true]{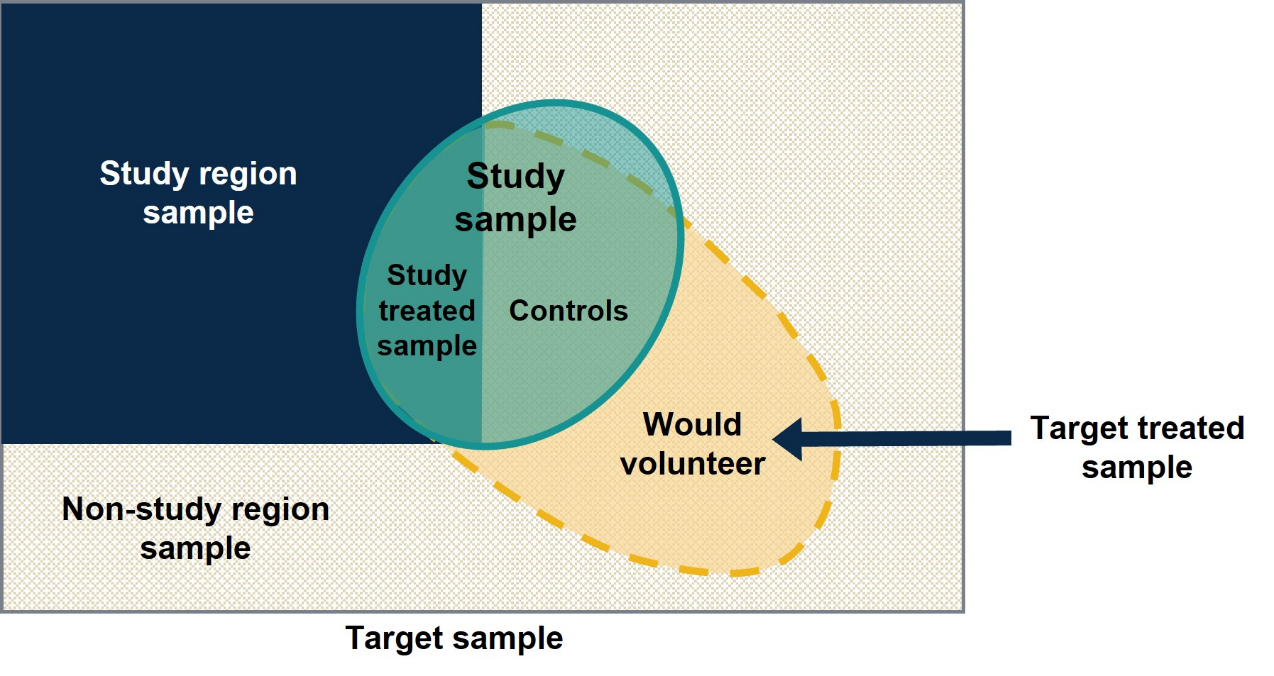}
		\caption{Volunteering in study and target samples and regions}
		\label{fig:DGP}		
	\end{center}
\end{figure}

We examine the setting of scaling up a voluntary intervention, as illustrated in Figure \ref{fig:DGP}, focusing on generalizing impacts of CPC+ nationwide. During the evaluation, CPC+ was offered within the study region. Eligible primary care practices in the study region that volunteered to participate in the intervention became part of the study treated population. Control primary care practices were then selected from neighboring regions using propensity score matching, though controls may be chosen in another manner in other studies. CPC+ practices and their matched control practices constitute the study sample. In the scale-up, the intervention would be offered to all nationwide primary care practices (the target population, across both study and non-study regions); those that volunteer to participate become part of the target treated population, to which we wish to generalize impacts of the intervention. The study region sample, study sample, target sample, and target treated sample are representative samples from their respective populations.

We observe $i = 1,\ldots,n$ independent and identically distributed copies of target sample practices' observed data, $O = (Y,A,\bm{X},S,R,V)\sim \mathcal{P} \in \mathcal{M}$, where $\mathcal{P}$ is the underlying probability distribution within $\mathcal{M}$, the collection of possible probability distributions, $Y\in \mathbb{R}$ is the outcome, $S \in \{0,1\}$ is an indicator for being in the study sample (treated or control), $R \in \{ 0,1\}$ is an indicator for being in the study region, $V \in \{ 0,1\}$ is an indicator for volunteering to participate in the scale-up, $A \in \{ 0,1\}$ is an indicator for being treated during the study (i.e., volunteering to participate and being enrolled in the study), and $\bm{X} \in \mathbb{R}^{d}$ is a vector of measured covariates. Capital letters denote random variables, while the realizations of these variables are denoted in lowercase letters (e.g., $a$, $\bm{x}$). Other observational units of interest are: data for the $i = 1,\ldots,n_{V = 1}$ target treated sample practices, $O_V = (Y,A,\textbf{X},S,R,V = 1)$; for the $i = 1,\ldots,n_{R = 1}$ study region practices, $O_R= (Y,A,\textbf{X},S,R = 1,V)$; for the $i = 1,\ldots,n_{S = 1}$ study sample practices, $O_S= (Y,A,\textbf{X},S = 1,R,V)$; and for the $i = 1,\ldots,n_{S = 1,A = 1}$ study treated sample practices, $O_A= (Y,A=1,\textbf{X},S = 1,R,V)$. We assume members of the study treated sample will not necessarily volunteer for the scale-up (for $O_A, V \neq 1$ necessarily), though they likely have a high propensity to volunteer.

As per the potential outcomes framework, let $Y^1$ be the potential outcome corresponding to participating in the scale-up intervention and $Y^0$ be the potential outcome corresponding to not participating. The estimand of interest is the TATT: the treatment effect in the target treated population, that is, among target population units that would volunteer for the intervention: $E(Y^1 - Y^0 | V = 1)$. This quantity is distinct from the study population average treatment effect among the treated (SATT), the treatment effect in the study treated population: $E(Y^1 - Y^0 | S=1, A = 1)$.

\subsection{Assumptions}

Identifying the TATT from the data relies on standard internal validity assumptions, standard external validity assumptions pertaining to the target treated population of volunteers \citep{stuart2011, tipton2013a, degtiar2023}, and an additional assumption regarding volunteering.

\noindent \textbf{Internal validity assumptions}:

\begin{enumerate}
	\def\labelenumi{\arabic{enumi}.}
	\item \emph{Conditional mean exchangeability of treatment assignment}:
		$E(Y^1-Y^0|S=1,\bm{X})=E(Y^1-Y^0|S=1,A,\bm{X})$. This means the evaluation study was not subject to unmeasured confounding: we adjusted for all variables that risk inducing internal validity bias if not appropriately accounted for.
	\item \emph{Positivity of treatment assignment:}
	$P(A = 1 | S = 1,\bm{X} = \bm{x} ) < 1$ for	$\bm{x} \in \mathbb{R}_{S = 1,A = 1}^{d} \subseteq \mathbb{R}^{d}$,
	where $\mathbb{R}_{S = 1,A = 1}^{d}$ is the covariate support for the study treated population. The covariate support of the study treated population is contained within the covariate support of the study comparison population.
	\item \emph{Stable unit treatment value assumption (SUTVA) for treatment assignment:} if $A=a$ then $Y=Y^a$. Units do not impact each other's outcomes and hence there is not nor will there be an added benefit or detriment from being in the same region as an existing intervention participant.	Also, although participants may have individually made different changes as a result of their intervention participation, the intervention is well-defined for all units.
\end{enumerate}

\noindent \textbf{External validity assumptions}:

\begin{enumerate}
	\def\labelenumi{\arabic{enumi}.}
	\setcounter{enumi}{3}
	\item	\emph{Conditional mean exchangeability of sample selection:}
		$E(Y^1-Y^0|V=1,\bm{X})= E(Y^1-Y^0|S=1,\bm{X})$.
	There are no unmeasured effect modifiers related to study membership (i.e., all effect modifiers whose distribution differs between study regions and non-study regions are measured). Thus, new enrollees in the scale-up can be expected to benefit to a similar degree as current participants with similar measured characteristics.
	\item	\emph{Positivity of sample selection among volunteers:}
	$P(S = 1 | V = 1,\bm{X} = \bm{x} ) > 0$ for	$x \in {\mathbb{R}_{V}^{d}\mathbb{ \subseteq R}}^{d}$, where $\mathbb{R}_{V}^{d}$ is the covariate support for the target treated	population. The target treated population could have taken part in the current study had it been implemented in their geography.
	\item \emph{SUTVA for sample selection:} if $S=s$, and $A=a$ then $Y=Y^a$.
	Potential outcomes under scale-up are not a function of whether an observational unit was in the initial study or the number of units in the intervention. Nor will the scale-up intervention differ from the evaluation study intervention (implementation by unit will remain the same). Furthermore, the study treated population would see benefits similar to those they observed to date, and---if the intervention were terminated---these current study treated participants would revert to their pre-intervention outcomes.
\end{enumerate}

\noindent \textbf{Volunteering assumption}:

\begin{enumerate}
	\def\labelenumi{\arabic{enumi}.}
	\setcounter{enumi}{6}
	\item
	\emph{Equivalent drivers of volunteering}: $P(V=1|\bm{X})=P(V=1|R=1,\bm{X})$. Which target population units volunteer for the scale-up would be driven by measured characteristics in a similar way as to what drove study treated units to participate within study regions.
\end{enumerate}

\subsection{Identification of the Estimand}

Under the above assumptions, the estimand can be identified as follows:
$$E(Y^1-Y^0|V=1)={\{E_{\bm{X}}[w(\bm{X})]\}}^{-1}E_{\bm{X}}[w(\bm{X})\tau(\bm{X})]$$ 
where $w(\bm{X})=P(V=1|R=1,\bm{X})$, $\tau(\bm{X}) = E(Y|S=1,A=1,\bm{X}) - E(Y|S=1,A=0,\bm{X})$, and $E_{\bm{X}}$ denotes the expectation over the randomness of $\bm{X}$ in the target population. See Appendix \ref{appendix:proof} for the derivation. 

\section{Estimating Target Population Average Treatment Effects Among the Treated}\label{sec:estimation}
\subsection{Estimation}
The TATT can be estimated by reweighting the target sample to resemble the target treated sample: $\hat{E}(Y^1-Y^0|V=1)=1/M\sum^M_{m=1}\{{[\sum^n_{i=1}w^m(\bm{X}_i)]}^{-1}\sum^n_{i=1}w^m(\bm{X}_i)\tau^m(\bm{X}_i)\}$, with $M$ corresponding to the number of posterior draws, and $w^m$ and $\tau^m$ representing the $m$-th draw from the posterior distribution of $w$ and $\tau$, respectively. A frequentist approach would estimate $\hat{E}(Y^1-Y^0|V=1)$ = ${[\sum^n_{i=1}{\hat{w}(\bm{X}_i)}]}^{-1}\sum^n_{i=1}{\hat{w}(\bm{X}_i)\hat{\tau}(\bm{X}_i)}$, with hats, $\hat{\cdot}$, denoting point estimates of their respective quantities.

We can similarly estimate the target population conditional average treatment effects among the treated (TCATTs) as $\hat{E}(Y^1-Y^0|V=1,\bm{X}=\bm{x})=$

$1/M\sum^M_{m=1}\{{[\sum^n_{i=1}w^m(\bm{x}_i)]}^{-1}\sum^n_{i=1}w^m(\bm{x}_i)\tau^m(\bm{x}_i)\}$. 

Estimating the TATT requires estimating propensity for volunteering weights and treatment effects for all units in the target sample:

\begin{enumerate}
	\item \textit{Estimate propensity for volunteering weights, $w$:} Fit a propensity for volunteering regression in the study region, then use it to estimate a posterior distribution of propensities for all units in the target sample. 
	
	\item \textit{Estimate treatment effects, $\tau$:} Fit an outcome regression (or alternative estimation approach, as described in Section \ref{sec:alt_estimators}) to the study sample, then use it to estimate a posterior distribution of treatment effects for all units in the target sample. 
\end{enumerate}

The TATT estimate consists of the mean across posterior draws of propensity-for-volunteering weighted averages of all target sample
treatment effects. 

\subsection{Incorporating Uncertainty with Respect to Target Treated Sample Membership}

When estimating treatment effects using Bayesian models, to account for the uncertainty in estimated volunteer status when predicting target treated sample impacts, each posterior draw of $\tau$ is multiplied by a posterior draw of $w$. This sequential approach, though not fully Bayesian, is an unbiased and valid approach for propagating uncertainty in the propensity for volunteering \citep{zigler2014, mccandless2009}, given identifiability assumptions. Propensity scores for volunteering should therefore be estimated using a Bayesian model that produces the same number of posterior samples as produced for $\tau$. For a frequentist estimation approach, uncertainty around the propensity to volunteer can be incorporated using a bootstrap in which both treatment effects and volunteering weights are re-estimated.

\subsection{Bayesian Additive Regression Trees and Bayesian Causal Forests for Generalizability}
We recommend obtaining $\tau$ estimates with a flexible regression approach such as BART or BCF, which flexibly model the response surface. BART is a Bayesian nonparametric outcome regression \citep{chipman2010, hill2011, green2012, kern2016}, which models the outcomes as a sum of binary regression trees with additive error: $Y_i=f(x_i,a_i)+{\epsilon}_i$ with $f(x_i,a_i)=\sum^{n_\text{trees}}_{j=1}{g(x_i,a_i,T_j,M_j)}$; ${\epsilon}_i\sim N(0,{\sigma}^2)$; $T_j$ denotes the tree structure for tree $j,$ $M_j$ denotes the bottom node means of tree $j$, $n_{trees}$ is the number of trees, and ${\sigma}^2$ is the error variance. A prior is placed on the tree functions $g(\cdot)$ to constrain each tree to be small and on the $M_j$ to keep the leaf node means near zero \citep{hill2011}. However, this prior may create ``regularization-induced confounding'' by over-shrinking confounding effects \citep{hahn2018,hahn2020}.

To overcome the risk of regularization-induced confounding, BCF introduces an estimate of the propensity score for treatment assignment, ${\pi}_A(x_i)=P(A_i=1|x_i)$, as an additional covariate and reparametrizes \textit{f} to allow for separate priors to be placed on confounding and effect modification \citep{hahn2020}: $Y_i=\mu (x_i,{\hat\pi}_A(x_i))+\tau (x_i)a_i+{\epsilon}_i$. The function $\mu$ captures the relationship between the outcome and confounders, while $\tau (x_i)a_i$ captures the relationship between the outcome and effect modifiers such that $\tau$ is the treatment effect. Errors may be heteroskedastic \citep{thal2021}: ${\epsilon}_i\sim N(0,{\sigma_i}^2)$.

As with BART, BCF is insensitive to hyperparameter specifications, thus requiring little hyperparameter tuning (the default priors work well across a wide range of settings), and it allows for inference through Bayesian posterior sampling \citep{hahn2020}. BCF has outperformed other causal estimators at causal inference competitions, such as those held at the Atlantic Causal Inference Conference \citep{thal2023, dorie2019}. BCF is particularly well-suited to addressing generalizability as it identifies effect modifiers in a data-driven fashion (rather than relying on subjective judgments to estimate heterogeneous treatment effects), it allows for full confounding control by separately regularizing confounding and effect modification, and it enables incorporating additional sources of uncertainty in a Bayesian fashion, such as uncertainty with respect to the propensity for volunteering.

In simulations, we explored whether including an estimate of the propensity for being in the study, $\hat{\pi}_S=\hat{P}(S|X)$, as a covariate in the $\tau$ component of the BCF outcome regression would improve finite-sample performance in a similar manner as when including the propensity for treatment as a covariate in the $\mu$ function. Addition of $\hat{\pi}_S$ provides a one-dimensional summary of the association between effect modifiers and selection into the study; the conditional mean exchangeability of sample selection assumption requires that an effect modifier be associated with study membership in order to lead to bias. However, as a practical consideration, including $\hat{\pi}_S$ precludes reusing the same outcome regression already fitted for estimating study impacts. Furthermore, Hahn et al. 2020 found that including the propensity for treatment as an effect modifier can degrade mixing; in simulations, we explore whether including the propensity for being in the study as an effect modifier may likewise hinder mixing.

\subsection{Alternative Estimators for $\tau$}\label{sec:alt_estimators}

The $\tau$ component of the TATT functional can alternatively be
estimated through other generalizability approaches such as IPPW, AIPPW,
or alternative outcome regression estimators like parametric ordinary
least squares (OLS) regressions. IPPW estimates $\hat{\tau} (X_i)=Y_i \hat{w}_{sa}(S_i=1,A_i=1,X_i)-Y_i \hat{w}_{sa}(S_i=1,A_i=0,X_i)$ where weights can be normalized for stability: $\hat{w}_{sa}(S_i=s,A_i=a,X_i)=\hat{w}^*_{sa}(S_i=s,A_i=a,X_i)/\sum^n_{i=1}{\hat{w}^*_{sa}(S_i=s,A_i=a,X_i)}$, and where $\hat{w}^*_{sa}(s,a,X_i)=I(S_i=s,A_i=a)/[P(A_i=a|S_i=s,X_i)P(S_i=s|X_i)]$. Alternatively, the combined propensity weights, including propensity for
volunteering weights, can be normalized for stability: $\hat{E}(Y^{1} - Y^{0} | V = 1 ) = \lbrack \sum_{i = 1}^{n}{\hat{w}_{\text{all}}(a_i = 1,\bm{X}_i )} \rbrack^{- 1}\sum_{i = 1}^{n}Y_i  {\hat{w}}_{\text{all}}( {a_i = 1\bm{,X}}_i ) - \lbrack \sum_{i = 1}^{n}{{\hat{w}}_{\text{all}}(a_i = 0,\bm{X}_i )} \rbrack^{-1}\sum_{i = 1}^{n}{Y_i  {\hat{w}}_{\text{all}}({a_i = 0\bm{,X}}_i )}$ where ${\hat{w}}_{\text{all}}({a_i,\bm{X}}_i ) = I( S_i = 1,A_i = a )\hat{P}(V_i = 1|\bm{X}_i\bm{)}/\lbrack\hat{P}(A_i = a|S_i = 1,\bm{X}_i)P(S_i = 1|\bm{X}_i)\rbrack$.

AIPPW estimates $\tau (X_i)= \hat{w}_{sa}(S_i=1,A_i=1,X_i)(Y-\hat{E}(Y|S_i=1,A_i=1,X_i))- \hat{w}_{sa}(S_i=1,A_i=0,X_i)(Y-\hat{E}(Y|S_i=1,A_i=0,X_i)) +\hat{E}(Y|S_i=1,A_i=1,X_i)-\hat{E}(Y|S_i=1,A_i=0,X_i)$. 

To date, OLS, IPPW, and AIPPW estimators for generalizability have relied on parametric regressions, which depend on correct model
specification for at least one of the regressions \citep{flores2013, kern2016, rudolph2017, dahabreh2020}. Flexible regression approaches have demonstrated superior performance to those that rely on parametric assumptions for estimating study population treatment effects \citep{thal2023, dorie2019}. The simulation compares these parametric approaches, BART, and BCF for estimating TATTs.

\section{Simulation Study Based on Real Data}\label{sec:sim}
\subsection{Methods}
We conducted a simulation to assess the finite sample performance of our novel TATT estimator for extending inference from an observational study of a voluntary intervention to the target treated sample units that would volunteer to participate in a scale-up. Because estimating treatment effects for a population that is not enumerable (volunteering units) is a novel consideration to the generalizability literature that no existing estimators to our knowledge have addressed, all estimators we compare use our TATT estimation approach, substituting different estimator choices for estimating impacts \textit{$\tau$} and weights \textit{w}.

We compared the performance of flexible regressions for estimating \textit{$\tau$ }and \textit{w }to using parametric least-squares regressions. Specifically, we compared:

\begin{enumerate}
	\item A BCF outcome regression for \textit{$\tau$ }that fit a BART propensity model for \textit{w }(``BCF'')
	
	\item BCF including the propensity for study selection as an effect modifier with BART propensity models (``BCF-\textit{$\pi_S$}'')
	
	\item BART outcome and propensity models (``BART'')
	
	\item A linear outcome regression that included interactions of each \textit{X }with \textit{A }but no higher-order interactions, that is, between combinations of \textit{X }with \textit{A}; it estimated the propensity model for \textit{w }with logistic regression (``OLS'')
	
	\item A Bayesian hierarchical model version of the OLS models that placed shrinkage priors on all confounding and effect modification parameters (``BHM'')
	
	\item IPPW with logistic propensity models (``IPW'')
	
	\item AIPPW with all regressions estimated using linear and logistic models (``AIPW'')
\end{enumerate}

Further estimator details are provided in Appendix \ref{appendix:sim_methods}. We likewise compared these estimators' relative performance for estimating TATT versus SATT (see Appendix \ref{appendix:estimators} for a description of the SATT version of each estimator). ``BCF,'' ``IPW,'' and other estimator abbreviations therefore refer to the versions of the estimators used for both TATT and SATT estimation. 

Specifically, we assessed these estimators' performance for estimating the TATT and SATT of a hypothetical CPC+-like intervention that aims to reduce Medicare spending among beneficiaries served by intervention practices. We generated data to reflect the CPC+ evaluation study and nationwide primary care practices \citep{swankoski2022, laird2022} with respect to practice and beneficiary covariate distributions (such as beneficiary age and practice size, urbanicity, multispecialty status, and ownership by a larger parent organization), matched comparison group design, and the longitudinal and hierarchical structure of the data. We introduced confounding by measured and unmeasured covariates, and effect modification by measured covariates, many of which differed in distribution between CPC+ and nationwide practices. The nonlinear response surface reflects that we found meaningful nonlinearities in real-world evaluation data, though the specific magnitude of nonlinearity was not calibrated to the real-world data. Our approach was similar to the data generating processes (DGPs) described in \cite{thal2023} and \cite{lipman2022}. Specifically, we generated simulated primary care practice data from study regions (n=11,000 practices) and non-study regions (n=37,000 practices) using the conditional distributions $P(Y,A,\bm{X},S,R,V)= P(R)P(\bm{X}|R)P(V|\bm{X})P(A,S|R,V,\bm{X})P(Y|A,\bm{X})$ described in Appendix \ref{appendix:sim_DGP}. Each simulated data set consisted of approximately 1,000 study treated practices, 4,000 study control practices, and 3,320 practices volunteering from non-study regions, for a total of approximately 4,320 volunteering practices (the target treated sample) out of 48,000 nationwide practices (the target sample; approximately 9\% were projected to volunteer).

For estimation, regressions were fit using practice-level simulated data (i.e., using practice-level averages of the practices' Medicare beneficiaries' covariates and outcomes) as prior work has demonstrated aggregate modeling can perform as well as analysis of individual-level data when treatment assignment occurs at that aggregated level \citep{thal2023}. In CPC+, practices (not individual beneficiaries) decided whether to join the intervention. As a result, treatment status varies at the practice level, so there is no within-practice variation in treatment status. Therefore, confounding because of beneficiary characteristics must be captured by between-practice differences in those characteristics. This implies that a practice-level regression will be able to adjust for all confounding, assuming it adjusts for the correct aggregations of beneficiary characteristics. (We acknowledge that those may be difficult to correctly guess in practice.)

To allow for heteroskedastic errors due to varying practice sizes, we weighted outcome regressions by the number of beneficiaries in each practice. We used 1,600 replications to ensure Monte-Carlo standard errors of the bias less than 0.2, based on standard errors of estimates being less than 8. Inference was conducted at the $\alpha = 0.1$ level, the standard for Medicare evaluations, with empirical 90\% uncertainty bounds formed from posterior draws or bootstrap replications. Simulations were run in R using the packages bcf for BCF \citep{thal2021}, dbarts for BART \citep{dorie2021}, brms for BHM \citep{burkner2017}, and MatchIt for matching \citep{ho2011}. Estimator specifications are described in Appendix \ref{appendix:sim_methods}.

\subsection{Results}

As a result of differences in effect modifier distributions between study and non-study populations, the true SATT in the simulated population was an increase in spending of \$0.99 per beneficiary per month (PBPM; 90\% sampling variability, i.e., variability in the truth across simulation replications, -2.89 to 4.86), whereas the true TATT in the simulated population was a savings of \$12.18 PBPM (90\% sampling variability -13.89 to -10.47; Appendix \ref{appendix:sim_results}). Thus, the simulated intervention showed null impacts in the study but would be cost saving in the target treated population. On average across simulation replications, all estimators besides AIPW correctly estimated unfavorable findings in the study treated population and savings in the target treated population, though estimators using parametric models overestimated effects by approximately \$9 to \$13, and nonparametric models overestimated effects by approximately \$5 to \$7 (overestimated the costs in the study treated sample and underestimated the savings in the target treated sample).

\begin{figure}[!t]
	\spacingset{1}
	\begin{center}
		\includegraphics[width=0.8\textwidth]{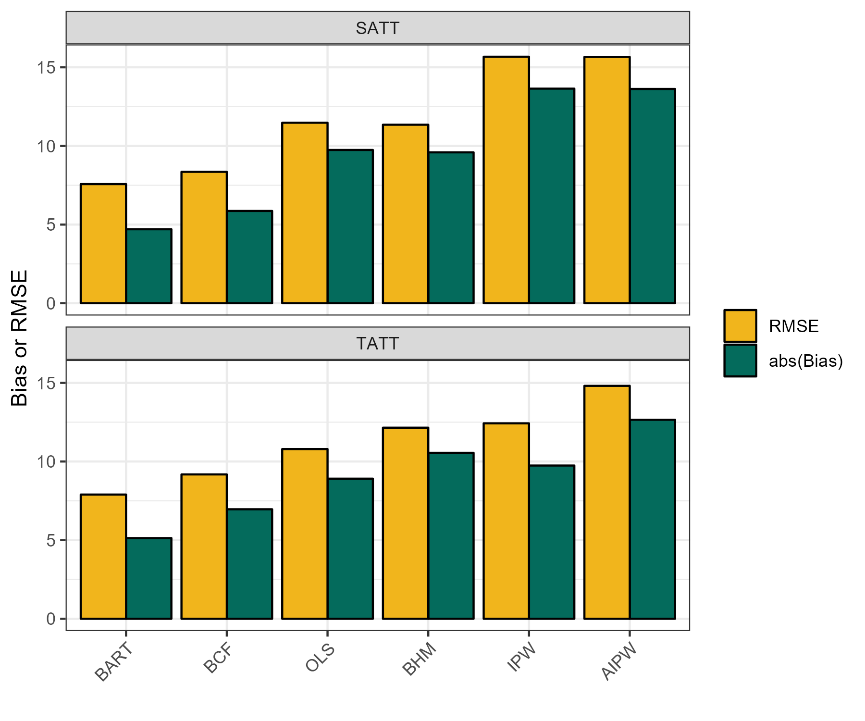}	
		\caption{Bias and RMSE for SATT and TATT estimates. Estimators are sorted in order of TATT RMSE. (A)IPW refers to (A)IPPW for TATT and (augmented) inverse probability of treatment weighting ([A]IPTW) for SATT. 
			Abbreviations: abs = absolute value; AIPW = augmented inverse probability weighting; BART = Bayesian Additive Regression Trees; BCF = Bayesian Causal Forest; BHM = Bayesian hierarchical model; IPW = inverse probability weighting; OLS = ordinary least squares; RMSE = root mean square error; SATT = study population average treatment effect among the treated; TATT = target population average treatment effect among the treated.}		
		\label{fig:results_bias_rmse}		
	\end{center}
\end{figure}

Nonparametric estimators (those fit with BART and BCF) exhibited smaller bias and RMSE than parametric approaches (those fit with linear and logistic regressions: OLS, BHM, IPW, and AIPW) for estimating the TATT and the SATT, as nonparametric estimators were better able to discover the complex confounding and effect heterogeneity relationships in the simulated data (Figure \ref{fig:results_bias_rmse}). For estimating the TATT under this DGP, BART performed best and AIPW performed worst, due to AIPW's combined misspecification of both outcome and propensity models. As also seen in prior work \citep{kang2007}, augmented inverse probability of treatment weighting (AIPTW) estimators' bias from misspecifying both outcome and propensity regressions may compound, potentially resulting in larger overall bias compared to a single misspecified outcome regression model. When using true rather than estimated impacts, AIPW (and IPW) performance exceeded that of other estimators (results not shown). IPW's large bias stemmed from the outcome's large variability and long tails: since IPW uses observed outcomes, it is subject to finite-sample bias when there are outliers in the outcome distribution, as also seen in previous work (e.g., from an outcome with long tails) \citep{canavire-bacarreza2021}. Overall, bias and RMSE were similar for estimating both TATT and SATT, for all estimators except IPW, whose bias for estimating TATT was surprisingly smaller than for estimating SATT. 

Across the three Bayesian estimators, BHM performed worse than BART and BCF. All three regularize parameter estimates, but BHM relies on linearity and additivity, while BART and BCF fit more flexible regressions, highlighting the importance of nonlinear functional forms, rather than solely regularization, to BART's and BCF's superior performance. However, all estimators exhibited residual bias, though the nonparametric estimators' bias was approximately half that of the parametric estimators.

\begin{figure}[!t]
	\spacingset{1}
	\begin{center}
		\includegraphics[width=0.8\textwidth]{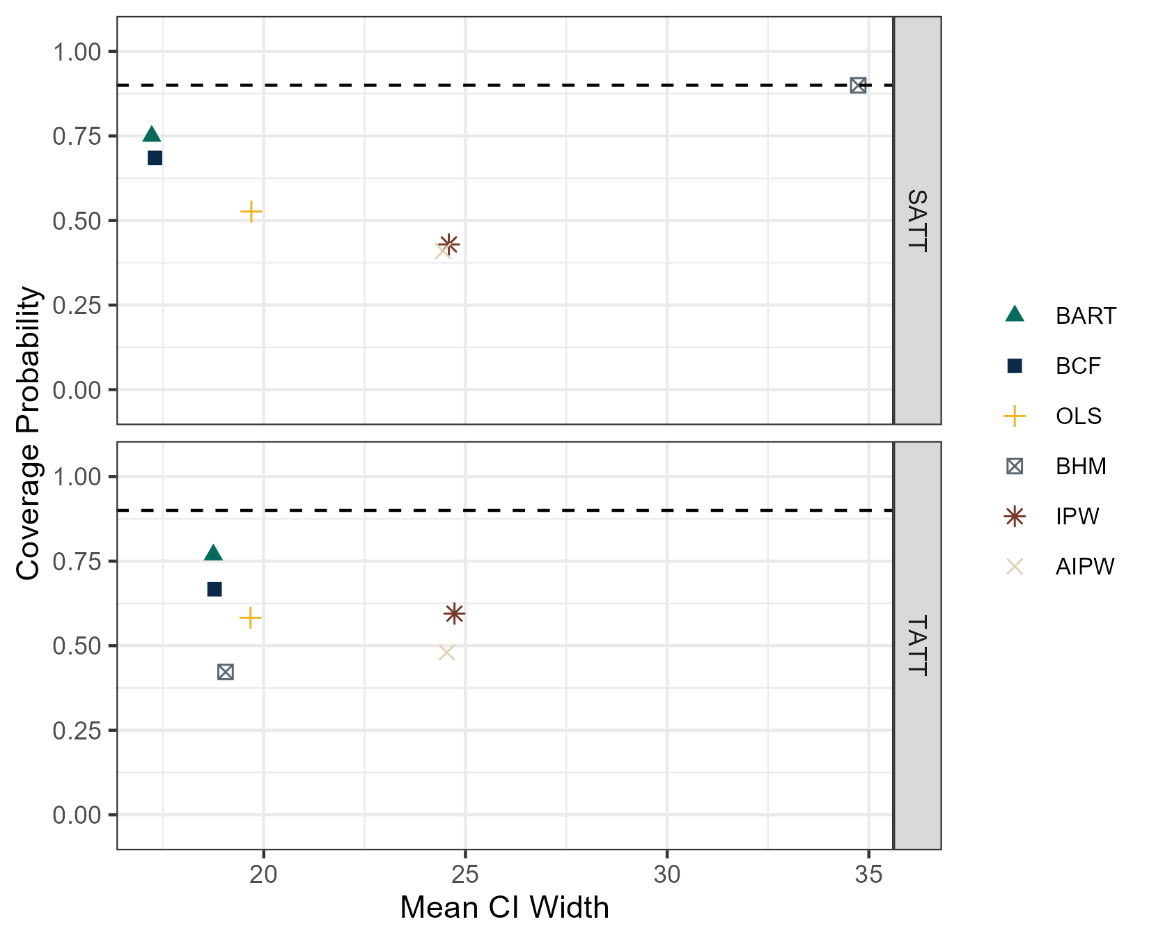}	
		\caption{Coverage and uncertainty bound width for SATT and TATT estimates. The dashed line corresponds to the target coverage of 90\%. (A)IPW refers to (A)IPPW for TATT and (augmented) inverse probability of treatment weighting ([A]IPTW) for SATT.
			Abbreviations: AIPW = augmented inverse probability weighting; BART = Bayesian Additive Regression Trees; BCF = Bayesian Causal Forest; BHM = Bayesian hierarchical model; CI = credible or confidence interval; IPW = inverse probability weighting; OLS = ordinary least squares.}		
		\label{fig:results_coverage}		
	\end{center}
\end{figure}

Residual bias led to undercoverage of both SATT and TATT for all estimators (except BHM for SATT; Figure \ref{fig:results_coverage}). TATT coverage was largely on par with SATT coverage. BART TATT coverage came closest to nominal---at 76\% for a target of 90\% coverage---while also having the smallest uncertainty interval width. Bias-adjusted coverage (i.e., the coverage of estimators after subtracting bias from the estimates) was just under nominal across all estimators (besides BHM, which overcovered for SATT, and BCF, which had nominal coverage for TATT; Appendix \ref{appendix:sim_results}), because estimated standard errors (SEs) just slightly underestimated empirical SE (except for BHM for SATT and BCF for TATT; Appendix \ref{appendix:sim_results}).

We had hypothesized that including $\hat{\pi}_S$ as an effect modifier in the BCF regression would improve finite-sample estimation performance but potentially degrade computational efficiency (Markov chain Monte Carlo mixing), based on corresponding observations made by Hahn et al. (2020) when including the propensity for treatment in BCF for estimating study average treatment effects. BCF-$\hat{\pi}_S$ did not noticeably improve performance over BCF for estimating TATT at the sample sizes in this DGP (bias and RMSE decreased slightly, by 0.1 and 0.06 respectively for the TATT, a difference smaller than the Monte-Carlo standard error). Including $\hat{\pi}_S$ did not seem to hinder mixing and, on average, it did not impact effective sample sizes or the potential scale reduction factor.

\section{Impacts of Scaling CPC+}\label{sec:cpcplus}
\subsection{ Methods}

\noindent Using our novel generalizability approach, we estimated what would have happened if, at the end of the CPC+ evaluation period, CMS had scaled up CPC+ instead of discontinuing the model. Although our estimates attempted (under identifiability assumptions) to forecast the future, they can be more precisely thought of as providing an accurate retrospective estimate of the impact if CPC+ been offered nationwide at the start of the evaluation. Specifically, this analysis examined, for the average beneficiary, the causal impact on total Medicare expenditures (without enhanced CPC+ payments) of scaling CPC+, as it was implemented in practices that joined CPC+ in 2017 through Program Year 4. (This analysis presents findings for scaling Track 1 of CPC+.) We estimated the effects of two scale-up approaches, corresponding to two target treated populations of interest: (1) a \textit{nationwide scale-up} and (2) a \textit{targeted scale-up} to practices in which the intervention would be likely to generate savings. The nationwide scale-up explored scaling up to all eligible practices nationwide that would volunteer, across both CPC+ regions and new regions. The targeted scale-up explored scaling up to practice types in which BCF estimated total Medicare savings from CPC+ to be most likely to offset care management fees---the cost of the intervention. 

To conduct the analysis, we used a nationwide data set of primary care practices (including CPC+ and comparison practices) and Medicare fee-for-service enrollment and claims data \citep{singh2020, swankoski2022}. First, we estimated which eligible nationwide primary care practices would likely volunteer for CPC+ by fitting a BART propensity score model for volunteering to participate among eligible practices in the 14 geographic regions that implemented CPC+ in 2017. Then, we used the fitted model to predict the propensity that each eligible practice nationwide---that is, across both CPC+ regions and non-CPC+ regions---would volunteer for a scale-up.

To estimate impacts, we fit a BCF outcome regression to the evaluation sample of CPC+ and matched comparison practices. Specifically, we fit practice-level models (rather than beneficiary-level models) to ensure computational tractability and because, as noted above, it is possible to adjust for all confounding with practice-level regressions in our setting (assuming correct aggregations of beneficiary characteristics). We used a version of BCF called individualized weighted BCF (iBCF), which allows larger practices to contribute more information to the likelihood than smaller practices through weights and includes practice-level random effects that allow for variability in the outcome and impacts that is not explained by covariates \citep{thal2021}. iBCF performed similarly to weighted BCF (without practice random effects) in our simulation study (Appendix \ref{appendix:sim_results}). Using the fitted iBCF regression, we predicted the impact of CPC+ among all nationwide practices. These impacts were then weighted by propensities to volunteer and by practice size to arrive at the TATT for the average beneficiary at a volunteering practice in a nationwide scale-up (i.e., $w(\bm{X})=n_i P(V=1|R=1,\bm{X})$). 

To identify subgroups expected to benefit most from the CPC+ intervention (which would therefore be promising candidates for a targeted scale-up), we used Classification and Regression Trees (CART) analysis of iBCF impact estimates \citep{hahn2020}. This CART fit-the-fit approach fit a heavily pruned CART tree (complexity parameter = 0.1) to the mean practice-level probabilities of offsetting care management fees (\$15 PBPM), estimated by iBCF. We focused on subgroups that improve CART's ability to explain impacts by at least 10\%. We then characterized the uncertainty for those subgroups via iBCF posterior draws. \cite{hahn2020} demonstrate that such an approach would have valid uncertainty quantifications. However, it may suffer from instability due to using a single tree, though various alternative specifications resulted in similar findings (results not shown). Further methodological details of the CPC+ evaluation study and scalability analysis can be found in \cite{degtiar2022}. 

\subsection{Results}

\begin{figure}
	\spacingset{1}
	\begin{center}
		\includegraphics[width=0.8\textwidth]{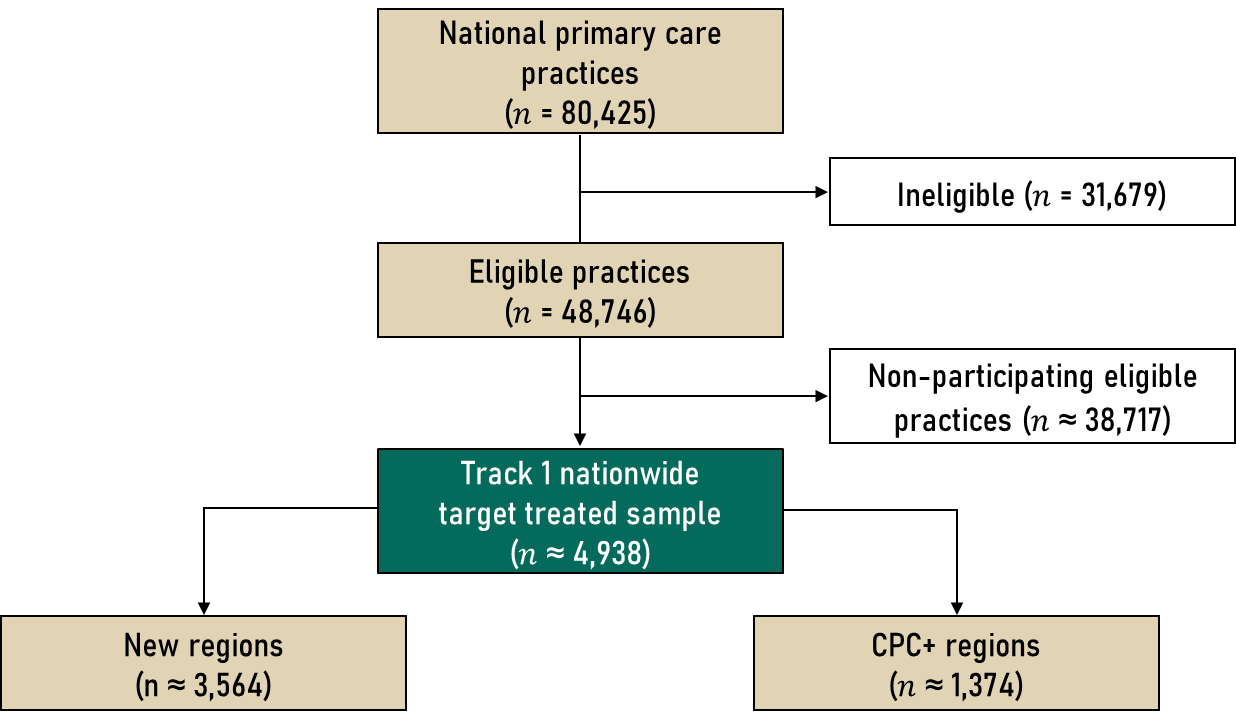}	
		\caption{Number of primary care practices in the United States, by eligibility and projected decision to volunteer for a nationwide CPC+ scale-up.}		
		\label{fig:cpcplus_consort}		
	\end{center}
\end{figure}

\begin{figure}
 	\spacingset{1}
 	\begin{center}
 		\includegraphics[width=0.8\textwidth]{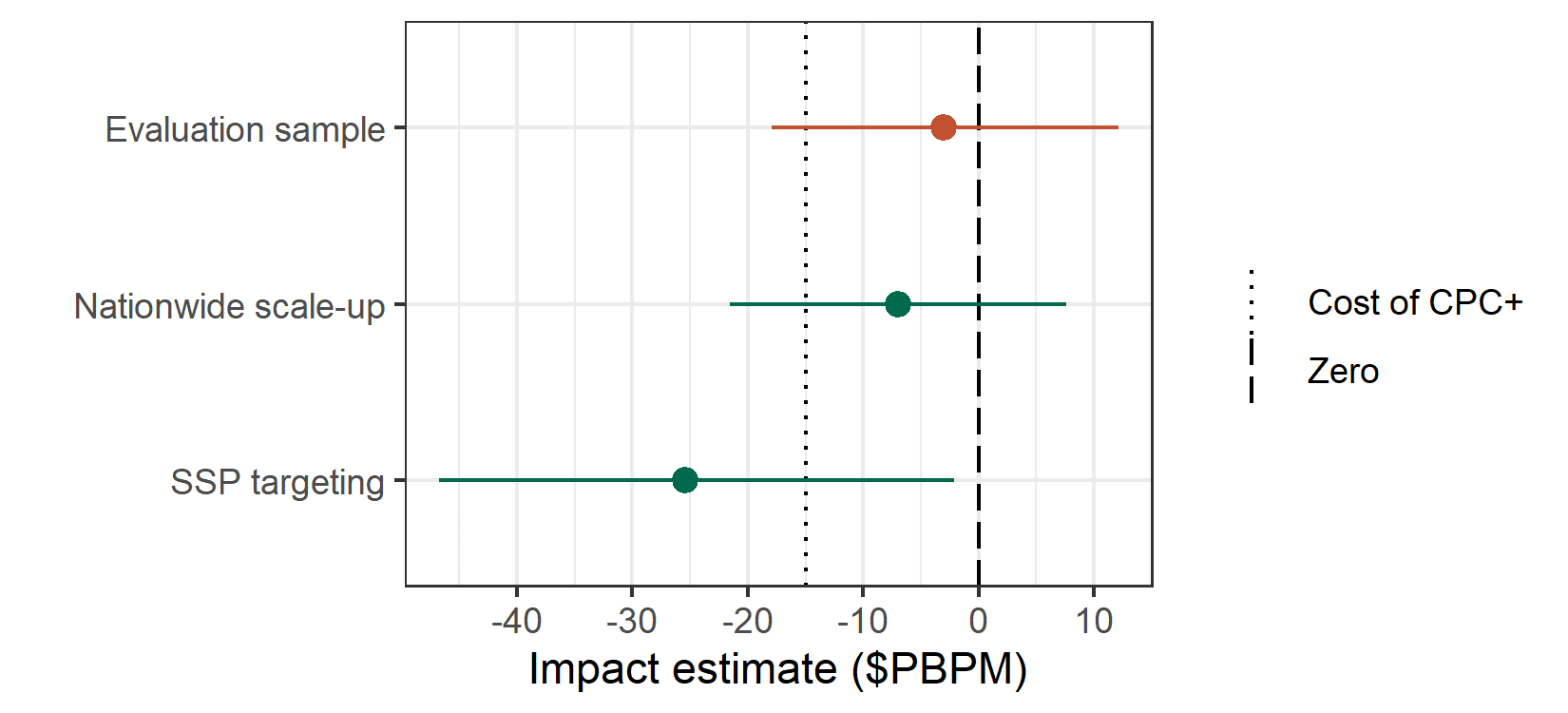}	
 		\caption{Estimated impacts of Track 1 of CPC+ on Medicare expenditures, excluding enhanced CPC+ payments in the evaluation sample and two target samples (estimate and 90\% CI. Abbreviations: CI = credible interval; PBPM = per beneficiary per month; SSP = Medicare Shared Savings Program.}		
 		\label{fig:cpcplus_results}		
 	\end{center}
 \end{figure}

 Of the 80,425 national primary care practices, we estimated 48,746 (61\%) would be eligible to participate and, of eligible practices, about 4,938 (10\%) would volunteer for Track 1 of CPC+ (Figure \ref{fig:cpcplus_consort}). While nationwide volunteers were estimated to differ from CPC+ participants on a range of characteristics (e.g., they had less experience with past primary care transformation), on the characteristic CART identified to be most strongly associated with favorable impacts---participation in the Medicare Shared Savings Program (SSP)---nationwide practices projected to volunteer were similar to CPC+ participants (54.6\% versus 53.8\% participation in SSP). Correspondingly, estimated impacts of a nationwide scale-up (-\$7 PBPM, 90\% credible interval [CI] -\$22 to \$8 PBPM) were similar to estimated impacts for CPC+ participants (-\$3 PBPM, 90\% CI -\$18 to \$12 PBPM; Figure \ref{fig:cpcplus_results}). However, targeting scale-up to SSP practices was estimated to have a 79\% probability of offsetting care management fees (compared to 18\% for a nationwide scale-up), for an estimated reduction of \$25 PBPM (90\% CI -\$47 to -\$2 PBPM). Thus, while a nationwide scale-up of Track 1 is unlikely to result in savings, a scale-up targeted to SSP practices (based on the definition of SSP at study baseline) might provide a path forward to thinking about achieving cost-neutrality in future primary care models. 
 
\section{Discussion}\label{sec:discussion}
 The proposed TATT estimation approach allows for estimating treatment effects among target treated population units that are not enumerable: those that would volunteer for the intervention in a scale-up. Using a BCF-based version of this TATT estimator, we projected CPC+ impacts from model participants to broader, policy-relevant populations. These projections, to all possible future volunteers and to the subgroup most likely to achieve cost-neutrality, can inform future policy models and model expansions, even when an intervention is estimated to be ineffective in the evaluation study overall. 

Estimating TATTs for a voluntary intervention, such as CPC+, is a novel contribution to the generalizability literature that no existing estimator addresses to our knowledge. Our approach does so via a weighted average of treatment effect estimates for all units in the target population to which the intervention would be offered, weighted by the propensity score for volunteering to participate. Posterior credible intervals reflect both uncertainty in treatment effect estimates across target sample practices and uncertainty in target treated sample membership.

In simulations we demonstrated that flexible outcome regression approaches such as BCF and BART improve performance over OLS, IPPW, and AIPPW estimators that rely on parametric regressions. BCF and BART estimators not only flexibly adjust for confounding (to ensure internal validity) but also flexibly model effect modification (to ensure external validity). In contrast, parametric regressions could not fully account for the complexity of the simulated response surface and so showed almost twice the bias of nonparametric estimators for both in-sample SATT estimates and out-of-sample TATT estimates. Thus, we demonstrated that flexible regressions can improve performance not just for estimating study population causal effects, as has previously been shown in American Causal Inference Conference data analysis competitions \citep{thal2023, dorie2019}, but also for estimating TATTs. Such prior work has likewise demonstrated that, in data generated to approximate the complexities of real-world policy evaluations, even flexible regressions are unable to fully debias estimates \citep{thal2023}, as also seen in the current simulation study for BART and BCF estimators. While BART outperformed BCF for estimating both SATTs and TATTs with our DGP, across other settings, such as with strong confounding and small effect sizes, BCF has demonstrated superior performance to BART \citep{hahn2020}. In addition to BART and BCF, other flexible approaches for discovering heterogeneous treatment effects could also be used, such as causal forests \citep{athey2016, wager2018}, neural network-based approaches \citep{johansson2018, shalit2017}, Gaussian process-based approaches \citep{alaa2017, alaa2018}, and ensembles \citep{grimmer2017, lee2020}. In contrast to these estimators, BART and BCF offer uncertainty bounds based on the posterior, ability to easily incorporate other sources of uncertainty due to being Bayesian estimators, and insensitivity to hyperparameter tuning \citep{hahn2020}.

A limitation of many flexible modeling approaches like BCF and BART is their computational burden. BCF is currently impractical to fit to data involving millions of observations, rather than thousands, and thus is unworkable for patient-level analyses of health policy interventions in lieu of the practice-level analyses we conducted in our simulations. For practice-level interventions such as CPC+, prior work has demonstrated that aggregation does not reduce performance \citep{thal2023}. However, even with practice-level data, BCF took approximately 30 minutes to fit and BART took approximately 3 minutes, while linear regressions took seconds (albeit their bootstrap took around 11 minutes, on an 11th Gen Intel(R) Core(TM) i7-1185G7 @ 3.00GHz 1.80 GHz processor). Ongoing work is improving the computational efficiency of these methods, including warm-start BCF \citep{hahn2020}, single program multiple data parallel computation for BART \citep{pratola2014}, and XBART \citep{he2019}. BCF's practicality can be further enhanced by extensions to available software to allow for correlated data analysis \citep{yeager2019}.

A further limitation is that estimators presented in this paper assess the impact of scaling up an intervention such as a policy model as it was offered in the study. The estimators therefore do not capture changes to the intervention under scale-up, changes to the setting such as the political landscape, nor changes to target treated population characteristics over time. Instead, although our estimates attempt to forecast the future, they can be more precisely thought of as providing an accurate \textit{retrospective} estimate of the impact if CPC+ had been offered nationwide in 2017. The estimators furthermore rely on identifiability assumptions that preclude unmeasured confounding and unmeasured effect modification, spillover effects, and different drivers of participation between study and non-study regions. To the extent that a prior distribution can be placed on bias resulting from assumption violations, there is room to incorporate these sources of uncertainty into the credible intervals. Otherwise, sensitivity analyses can assess the impact of such assumption violations on TATT estimates. 

Furthermore, for assessing impacts of scaling CPC+, our estimator focused on one year of evaluation findings, from Program Year 4. Future work should consider flexible longitudinal regressions to capture year-to-year heterogeneity in impacts, to avoid overinterpreting impacts from any one year of an evaluation. Despite these limitations, the current work takes a step towards better informing an evidence-based scale-up by accounting for differences in characteristics between study treated and anticipated target treated populations.

As demonstrated in our CPC+ work, the generalizability estimator presented here can also be applied to identify alternative feasible target populations for a scale-up, such as populations expected to benefit most from the intervention, defined by values of key effect modifiers. Estimates from such targeting approaches can inform future policy models and model expansions, particularly when a broader scale-up is estimated to be ineffective. By estimating treatment effects for the target treated population of interest, our generalizability estimator provides impacts in a policy-relevant population to better guide future policy decisions.

\if1\blind
{
	\section{Acknowledgments}
	We thank Tim Day, Dan Thal, Laura Blue, John Deke, Nancy McCall, and Sherri Rose for helpful input provided on the simulation and CPC+ work. We thank the CPC+ evaluation team at Mathematica for the enormous data processing work that enabled the CPC+ study.   

	\section{Conflicts of Interest}
	The authors report there are no competing interests to declare.
	
} \fi

\bibliographystyle{Chicago}
\bibliography{Library_PATT_estimation.bib}

\appendix
\section*{Appendix}
\renewcommand{\thetable}{A\arabic{table}}
\renewcommand{\thefigure}{A\arabic{figure}}
\renewcommand{\thesection}{A\arabic{section}}
\section{Proof for Identification of $E(Y^1-Y^0|V=1)$}\label{appendix:proof}

{\footnotesize 
	\begin{align}
		E(Y^1-Y^0|V=1) & = E_{\bm{X}}\big[E(Y^1|V=1,\bm{X})-E(Y^0|V=1,\bm{X})\big|V=1\big] \\
		& =E_{\bm{X}}\big[E(Y^1|S=1,\bm{X})-E(Y^0|S=1,\bm{X})\big|V=1\big] \\
		& =E_{\bm{X}}\big[E(Y^1|S=1,A=1,\bm{X})-E(Y^0|S=1,A=0,\bm{X})\big|V=1\big] \\
		& =\frac{1}{P(V=1)}E_{\bm{X}}\big[V\{E(Y|S=1,A=1,\bm{X})-E(Y|S=1,A=0,\bm{X})\}\big] \\
		& =\frac{1}{E_{\bm{X}}[P(V=1|\bm{X})]}E_{\bm{X}}\big[P(V=1|\bm{X})\{E(Y|S=1,A=1,\bm{X})-E(Y|S=1,A=0,\bm{X})\}\big] \\
		& =\frac{1}{E_{\bm{X}}[P(V=1|R=1,\bm{X})]}E_{\bm{X}}\big[P(V=1|R=1,\bm{X})\{E(Y|S=1,A=1,\bm{X})-E(Y|S=1,A=0,\bm{X})\}\big] \\
		& ={E_{\bm{X}}[w(\bm{X})]\}}^{-1}E_{\bm{X}}[w(\bm{X})\tau (\bm{X})] 
	\end{align}
}%
Line 1 follows from the law of total probability, line 2 from conditional sampling exchangeability, line 3 from conditional treatment exchangeability, line 4 from Bayes rule and SUTVA for treatment assignment and sample selection, line 5 from the law of total probability, and line 6 from equivalent drivers of volunteering. Positivity of treatment assignment and sample selection among volunteers are needed for the functionals to be well-defined.

\section{Simulation Estimator Details}\label{appendix:sim_methods}
For outcome regressions (and propensity regressions, for BART and BHM estimators), each BCF run used 3 chains with 300 posterior samples each after discarding 3,000 samples as burn-in (500 for iBCF) and thinning by 30 (900 total retained posterior samples); each BART run used 900 posterior samples after thinning by 30 (by 3 for propensity regressions) and discarding a default of 100 burn-in; each BHM run used 3 chains with 300 posterior samples each with no thinning and a default of 450 burn-in (900 total retained posterior samples); frequentist inference was based on 900 bootstrap draws. BCF and BART used default priors. The BHM outcome regression likelihood was specified as follows:
$$y^*_i \sim \mu_0 +\mu + \tau_0 A_i + \bm{\tau}\bm{X}^*_i A_i + \epsilon_i$$ 
where $(\cdot)^*$ indicates a centered variable scaled to have a standard deviation of 1. $\bm{X}^*$ consisted of 2 continuous covariates and indicators for 3 binary and 2 categorical covariates, such that $\mathrm{dim}(\bm{X}^*)=n \times 14$. The priors on model parameters were as follows:
\begin{align*}
	\mu_0 &\sim N(0,1) \\
	\tau_0 &\sim N(0,1) \\
	\mu &\sim N(0,\sigma_{\mu}^2)\ \mathrm{ where }\ \sigma_{\mu} \sim N^+(0,1) \\
	\tau &\sim N(0,\sigma_{\tau}^2)\ \mathrm{ where }\ \sigma_{\tau} \sim N^+(0,1) \\
	\epsilon &\sim N(0,\nu^2/n_{\mathrm{bene}})\ \mathrm{ where }\ \nu \sim N^+(0,1)
\end{align*}

\noindent and $n_{\mathrm{bene}}$ was the number of beneficiaries in the practice. We included a sum-to-zero constraint on the coefficients for each of the binary/categorical variables.

The propensity for volunteering likelihood had a parallel specification: $$logit(P(V_i=1)) \sim \beta_0 +\bm{\beta}\bm{X}_i^{**}$$ where $(\cdot)^{**}$ indicates a centered variable scaled to have a standard deviation of 0.5. We specified the following priors:
\begin{align*}
	\beta_0 &\sim t(3,0,2.5) \\
	\bm{\beta} &\sim N(0,\sigma_{\beta}^2)\ \mathrm{ where }\ \sigma_{\tau} \sim N^+(0,2.5)
\end{align*}

\section{SATT Estimators}\label{appendix:estimators}

\noindent The SATT, $E(Y^1-Y^0|S=1,A=1)$, was estimated as follows:

\begin{enumerate}
	\item OLS: $1/n_{S=1,A=1}\sum^{n_{S=1,A=1}}_{i=1}{{\tau (\bm{X}_i)} I(S_i = 1,A_i = 1)}$
	
	\item Inverse probability of treatment weighting (IPTW):
	$$\sum^{n_{S=1}}_{i=1}{Y_i  w_a(S_i = 1,A_i = 1,\bm{X}_i)-Y_i w_a(S_i=1,A_i=0,\bm{X}_i)}$$ 
\end{enumerate}
where weights were normalized for stability, $w_a(S_i=s, A_i=a, \textbf{X}_i) = w_a^*(S_i= s,A_i= a, \textbf{X}_i) / \sum^n_{i=1}{{w_a}^*(S_i=s,A_i = a, \textbf{X}_i})$, with ${w_a}^*(S_i=1,A_i=1,\bm{X}_i) = I(S_i=1,A_i=1)$ and
$$w^*_a(S_i=1,A_i=0,\bm{X}_i)=I(S_i=1,A_i=0)\frac{P(A_i=1| S_i=1,\bm{X}_i)}{1-P(A_i=1|S_i=1,\bm{X}_i)}$$ 

\begin{enumerate}
	\item Augmented inverse probability of treatment weighting (AIPTW):
	\begin{align*}
		&\frac{1}{n_{S=1,A=1}}\sum^{n_{S=1}}_{i=1}\Big\{Y_i  I(S_i=1,A_i=1)-\big\{Y_i  I(S_i=1,A_i=0)  w_a(S_i=1,A_i=1,\bm{X}_i)\big\} + \\ 
		&\hat{E}(Y|S=1,A=0,\bm{X}_i) \big[I(S_i=1,A_i=1)-w_a(S_i=1,A_i=1,\bm{X}_i)\big]\Big\}/\big[1-w_a(S_i=1,A_i=1,\bm{X}_i)\big]
	\end{align*}
	
\end{enumerate}

\section{Simulation Data Generating Process}\label{appendix:sim_DGP}
For the simulation, we generated $P(Y,A,\bm{X},S,R,V)=P(R)P(\bm{X}|R)P(V|\bm{X})P(A,S|R,V,\bm{X})P(Y|A,\bm{X})$ as follows:

\begin{enumerate}
	\item $R$ and $X_1,\dots,X_7$: The real data on which the simulation was based consisted of beneficiary- and practice-level claims data for the CPC+ evaluation study and for all nationwide practices eligible for the scaled-up intervention. As most baseline covariates were categorical, we simulated baseline beneficiary- and practice-level covariates for the study region and non-study regions separately in the same proportions observed in the real data based on nonparametric assignment of practices to combinatoric cells describing possible baseline characteristic combinations (similar to that described in \cite{lipman2022}). We then aggregated beneficiary-level data to the practice level by taking practice-level means of beneficiary characteristics. 
	
	\item $V|\bm{X}\sim binom(expit(\bm{X \gamma}))$, with $binom$ corresponding to the binomial distribution and $expit(P)=exp(P)/[1+exp(P)]$: We fit a logistic propensity for volunteering regression to CPC+ study region practices (based on proxies, $X_2$, $X_4$ and $X_5$) for factors that drove volunteering in CPC+) to obtain $\bm{\gamma}$'s, which we used to generate a propensity for volunteering process for all nationwide practices: $P(V_j=1|\bm{X}_j)=-5.23+1.25B_j+(1-B_j)(1.25I(X_{4,j}=1,X_{5,j}=1)-1.25I(X_{4,j}\neq 1|X_{5,j}\neq 1)+0.04X_{2,j}$, with $B_j\sim binom(0.25)$. Volunteering status was generated for each practice with probability corresponding to its propensity. Approximately 9\% of practices volunteered, as was observed in the real data.
	
	\item $A$ and $S$: Within the study region, practices that volunteered became the study treated group. To parallel the CPC+ control selection process, study controls were then chosen from non-study regions via 4:1 matching. Controls were matched to study treated patients based on their propensity scores for treatment (estimated via logistic regression) using the MatchIt package in R \citep{ho2011}. Study region volunteering practices and matched controls from the non-study region formed the study sample. 
	
	\item $Y|A,\bm{X}$: We used an outcome generating process in which beneficiary sickness levels drove practice spending. We generated the outcomes at the beneficiary level as $Y_{ijt}=\mu(\bm{X}_j)+c_0-45t+c_1 U_{1,j}+c_2U^2_{1,j}+\tau(\bm{X}_{ij},\nu_i)A_j+ \epsilon_i$, with $i$ indexing beneficiaries and $j$ indexing practices, $t\in \{0,1\}$ indicating pre/post-intervention time, the outcome of interest being $Y=Y_{t=1}-Y_{t=0}$; $U_{1,j}$ corresponding to practice-level averages of beneficiary sickness levels, determined through a stochastic process described in Lipman et al. (2022) ($U_{1,j}$ is an unmeasured confounder proxied by the measured variable beneficiary age=$X_{1,i}$ and its practice-level mean $X_{2,j}$); $\mu(\bm{X}_j)$ corresponding to additional confounding, namely $\mu(\bm{X}_j)=-128X_{3,j}X_{9,j}+48\mathrm{log}(X_{2,j}+0.1) +96X_{5,j}X_{6,j}-64X_{8,j}$; the treatment effect function consisting of effect heterogeneity driven by both measured practice-and-beneficiary-level ($\bm{X}_{ij}$) and unmeasured beneficiary-level ($\nu_i)$ variables is $\tau(\bm{X}_i,\nu_i)=-69+50\sqrt{0.2}  \nu_i+\sqrt{0.8} \lambda (\bm{X}_{ij})$, with $\nu_i\sim N(0,1)$, and $\lambda (\bm{X}_{ij})=\sqrt{0.3}(-max(0,{\mathrm{log} (max(X_{1,i}-70, 0))})+ X_{3,j}(I[X_{1,i}<65]+I[X_{1,i}<72])+ \bm{X}^T_j\bm{\beta }+(\bm{X}_j:\bm{X}_j)^{T}\bm{\delta}, \bm{X}_j=[X_{3,j},\dots,X_{9,j},t]$, where $\bm{X}_j:\bm{X}_j$ is the matrix of interaction terms between each of the covariates $\bm{X}_j$, $\bm{\beta}=[-0.42\sqrt{0.5}, 1.72, -0.45, 0.34,,-0.49,-0.17, 1.80, 1.39],\bm{\delta}=\sqrt{0.05}  [1.81, 0.10, 0.33, 0.93, $ $-2.21, 2.78, -0.66, -0.04, -0.53, 0.88, 1.66, -0.01, 0.24, -0.26, 1.24, 0.71, 1.08,-0.78,$ \newline 
	$ 0.03, 1.73, -0.67, 0.34, 1.15, -0.06, -0.74, 0.29], \epsilon_i= max(25logN(0,2.2),100000)$ where $logN(\cdot)$ corresponds to the log normal distribution; $c_0, c_1, c_2$ were unmeasured practice-level cost-multipliers that correspond to different practices incurring different costs for treating the same sickness levels \citep{lipman2022}. After generating $Y_{ijt}$, we added an offset to negative values to ensure all spending exceeded 0.
\end{enumerate}

\section{Simulation Results}\label{appendix:sim_results}

\scriptsize
\spacingset{1}
\begin{tabular}{p{0.5in} p{0.5in} p{0.3in} p{0.3in} p{0.3in} p{0.4in} p{0.5in} p{0.5in} p{0.5in} p{0.5in} p{0.6in} } \hline
	\scriptsize{\textbf{Estimand}} & \scriptsize{\textbf{Estimator}} & \scriptsize\textbf{Bias} & \scriptsize\textbf{SE} & \scriptsize\textbf{RMSE} & \scriptsize\textbf{CI width} & \scriptsize\textbf{Coverage} & \scriptsize\textbf{Bias-adjusted coverage} & \scriptsize\textbf{Estimate, mean} & \scriptsize\textbf{Estimated SE, mean} & \scriptsize\textbf{Probability of savings} \\\hline 
	SATT & BART & 4.70 & 5.94 & 7.57 & 17.22 & 0.76 & 0.86 & 5.69 & 5.25 & 0.24 \\
	SATT & BCF-${\pi }_S$ & 5.77 & 5.95 & 8.29 & 17.30 & 0.71 & 0.85 & 6.75 & 5.27 & 0.20 \\
	SATT & BCF & 5.87 & 5.95 & 8.35 & 17.30 & 0.69 & 0.86 & 6.85 & 5.28 & 0.19 \\
	SATT & iBCF & 6.02 & 5.99 & 8.49 & 17.83 & 0.70 & 0.86 & 7.00 & 5.44 & 0.19 \\
	SATT & OLS & 9.75 & 6.06 & 11.48 & 19.69 & 0.51 & 0.89 & 10.73 & 6.01 & NA \\
	SATT & BHM & 9.59 & 6.07 & 11.35 & 34.74 & 0.91 & 1.00 & 10.58 & 10.60 & 0.19 \\
	SATT & IPW & 13.64 & 7.72 & 15.67 & 24.60 & 0.42 & 0.89 & 14.63 & 7.50 & NA \\
	SATT & AIPW & 13.62 & 7.72 & 15.66 & 24.43 & 0.42 & 0.89 & 14.61 & 7.45 & NA \\
	TATT & BART & 5.13 & 6.01 & 7.90 & 18.76 & 0.76 & 0.88 & -7.05 & 5.72 & 0.80 \\
	TATT & BCF-${\pi }_S$ & 6.79 & 5.99 & 9.06 & 18.80 & 0.67 & 0.89 & -5.39 & 5.73 & 0.74 \\
	TATT & BCF & 6.96 & 5.98 & 9.18 & 18.77 & 0.66 & 0.88 & -5.22 & 5.73 & 0.74 \\
	TATT & iBCF & 7.11 & 6.02 & 9.32 & 19.51 & 0.66 & 0.90 & -5.07 & 5.95 & 0.73 \\
	TATT & OLS & 8.91 & 6.09 & 10.79 & 19.69 & 0.57 & 0.89 & -3.27 & 6.00 & NA \\
	TATT & BHM & 10.55 & 6.03 & 12.15 & 19.04 & 0.44 & 0.89 & -1.63 & 5.80 & 0.58 \\
	TATT & IPW & 9.74 & 7.72 & 12.43 & 24.74 & 0.61 & 0.89 & -2.44 & 7.55 & NA \\
	TATT & AIPW & 12.66 & 7.70 & 14.81 & 24.52 & 0.48 & 0.88 & 0.48 & 7.48 & NA \\\hline 
	\multicolumn{11}{p{\textwidth}}{\textit{Abbreviations: AIPW = augmented inverse probability weighting; BART = Bayesian Additive Regression Trees; BCF = Bayesian Causal Forest; BHM = Bayesian hierarchical model; CI = credible or confidence interval; IPW = inverse probability weighting; NA = not applicable; OLS = ordinary least squares; RMSE = root mean squared error; SATT = study population average treatment effect among the treated; SE = standard error; TATT = target population average treatment effect among the treated.}}
\end{tabular}
\end{document}